\documentclass[longbibliography,superscriptaddress]{revtex4-2}

\usepackage[utf8]{inputenc}
\usepackage{amsmath}
\usepackage{amssymb}
\usepackage{amsmath}
\usepackage{appendix}
\usepackage{bbold}

\begin{document}
\title{    Ramsey Scheme  Applied to String Theoretical Processes}

\author{Salman Sajad Wani}
	\affiliation{Department of Physics and Engineering, Istanbul, Technical University, 34469, Istanbul, Turkey}
\affiliation{Canadian Quantum Research Center 204-3002, 32 Ave Vernon, BC V1T 2L7 Canada}

\author{Arshid Shabir }
\affiliation{Department of Physics, NIT Srinagar, Kashmir 190006 India}

\author{Mir Faizal}
\affiliation{Canadian Quantum Research Center 204-3002, 32 Ave Vernon, BC V1T 2L7 Canada}
\affiliation{Department of Physics and Astronomy, University of Lethbridge, Lethbridge, AB T1K 3M4, Canada}

\author{Seemin Rubab}
\affiliation{Department of Physics, NIT Srinagar, Kashmir 190006 India}

\begin{abstract}
In this letter, we analyze the evolution of physical quantities due to the interaction of strings with background fields. We will obtain the characteristic function for such a string theoretical process. This will be done by generalizing the Ramsey scheme to world-sheet, and using it to obtain the information about the evolution of quantity in a string theoretical process, without making two-point measurements.  We will also use the characteristic function to obtain the average of the difference between the initial and final values of such a quantity.  Finally, using the characteristic function,  we calculate fisher information for the difference of such a quantity.   
\end{abstract}

\maketitle
\section{Introduction}
It may be noted that  ordinary particles are  described by a number of quantities, like mass and spin,  and these quantities are not dynamically obtained in quantum field theories. However,  in string theory, these can be obtained from the dynamics of world-sheet \cite{x1, x2}. This is because  strings have an extended structure, and have their own internal dynamics associated with  that internal structure. So,  unlike particles in ordinary quantum field theory (which have no internal dynamics and are just points in space-time), the internal dynamics of a world-sheet of string theory are represented by  a  conformal  field theory. Thus, such quantities in ordinary quantum field theory can be dynamically calculated using this conformal field theory describing the world-sheet dynamics of string theory. In string theory, there are background fields. These fields can interact with the strings, and change their oscillatory modes \cite{y0,y4,   y1, y2}. The change in the oscillatory modes would also change the   value of quantities, such as mass and spin, which can be obtained dynamically from the world-sheet dynamics. It is important to analyze the change in the value of such quantities due to the interaction of strings with background fields.  

It is known that the change in a quantity  due to the interaction of a system with some external potential can be obtained using two-point measurement   scheme \cite{tasaki00,kurchan01}.  In the  two-point measurement  scheme, a system interacts with a field, and this changes the initial eigenvalues of an operator to  final eigenvalues. The difference between these  
two projective measurements of a quantity,   made at  the beginning and end of such a  process, are then used to obtain the distribution function for the quantity \cite{tasaki00,kurchan01}.
This difference is the difference between the initial and final eigenvalues of the operator, after the system has interacted with an external field. 
Such  two projective measurements cannot be  made on a system with an intrinsic Lorentz structure, due to problems like violation of  locality and superluminal signals \cite{5cd,5mn, 6lk}. However, these problems can be resolved by using  the Ramsey scheme  \cite{6ab, 6ba}.  The Ramsey scheme has been applied to quantum field theory \cite{5}. 

In this scheme,  first an auxiliary qubit is coupled to the system, and then the   information about the system is transferred to the qubit. The information about the system is obtained by making   measurements on the qubit.  The  effects of the   evolution  of the system by its interaction with an external field can also be probed using this qubit. This is done by  preparing the qubit in a superposition of ground and excited states, and then  transferring  the information  about the characteristic function of the  distribution to the state of the qubit. This  procedure can be used to obtain the information about the change in the eigenvalues of an operator, without making projective measurements. The  Ramsey scheme  has been used to obtain such a difference in quantum field theory \cite{5}. As the dynamics of the world-sheet of string theory is represented by a conformal field theory, in this letter, we will generalize the Ramsey scheme to string theory.

We will also analyze the flow of information on the world-sheet of strings. This will be done by using quantum fisher information. In fact, quantum fisher information can be used to analyze the behavior of those quantities, which are not represented by a quantum   operator  \cite{1a, 2a, 1b, 2b}. As the difference of the initial and final values of a quantity,  in a string theoretical process, cannot be represented by a quantum operator, we will use quantum fisher information to analyze its behavior. This quantum fisher information would be obtained  from the distribution of the difference between the initial and final values of a given quantity.  Here we point out that even though the original quantity can be represented by a quantum operator, the change in the quantity in a quantum process cannot be represented by any quantum operator. This is similar to  the Hamiltonian being  a quantum operator, but quantum work (which represents the change in the eigenvalues of the Hamiltonian in a quantum process) cannot be represented by any operator  \cite{4a, 4b, 4c, 4d}.

\section{String Theoretical Process}
In this section, we will analyze string theoretical process in which a string interact with a background field. To analyze such a process, we will start from the free  Polyakov action, which describes  the world-sheet of bosonic  string theory 
\begin{eqnarray}
     {S} =   \frac{1}{4 \pi \alpha'} \int d\sigma ^2   \eta^{\alpha\beta} \partial_{\alpha}X_{\mu}  \partial_{\beta} X^{\mu}.
\end{eqnarray}
Here  $\alpha'$ is   related to the  length  of string $l_s^2$. We can apply the 
  Neumann boundary conditions for open strings, and write  
\begin{eqnarray}
\label{sol}
    X^{\mu}\left(  \tau,\sigma\right)=  x^{\mu}+l_s \tau p^\mu+ il_s\sum_{m\neq0}\frac{1}{m}{{\alpha}}^{\mu}_m e^{-im\tau}cos(m\sigma ).
    \end{eqnarray}

As string  theory has a gauge degrees of freedom, we need to fix a guage before  quantizing  it. This can be done using the BRST quantization in the conformal gauge \cite{gauge}. This is done by adding a gauge fixing term (conformal gauge) and a ghost term to the  Polyakov action. This new effective action obtained by the sum of the original Polyakov action with the gauge fixing and ghost term is invariant under BRST transformation. The physical states of the system are then constructed using the   Noether charge corresponding to this symmetry of the effective action. This Noether change is called the BRST charge, $Q$, and physical states satisfy $Q |\rm{phy}> =0$. However, as the BRST quantization in conformal guage is equivalent to light-cone quantization \cite{gauge1, gauge0}, we can use analyze string states in  light-cone \cite{gauge2, gauge4}. Furthermore, the   covariant string theory, 
splits into the light-cone string theory and trivial excitations   \cite{gauge5}. The other excitations which represent the ghost sector decouple from the theory, and can be neglected in light-cone quantization \cite{gauge12, gauge14}.  This is similar to the situation in ordinary electrodynamics, where the ghost sector decouples from the theory, and can be neglected for perturbative calculations 
\cite{guage15}. 
Thus, it is possible to analyze string theory in light-cone gauge \cite{gauge2, gauge4}. 
This finite field BRST transformation is not   the symmetry of the path integral
measure, and the   Jacobian of transformation can be used to transform the  generating functional from light-cone gauge to covariant gauge \cite{a1, a2, a3, a4,  a5,a6, a7, a8}. In fact, it has also been demonstrated that finite field dependent BRST transformation can be used to obtain the  generating functional in light-cone gauge from the  generating functional conformal gauge \cite{gauge8}.
Here again we observe that in the  light-cone gauge  the gauge degrees of freedom are removed, and so the string theory can be consistently quantized in it. This is done by first expressing  the target space  coordinates  as $\{ X^+, X^-,X^i\}_{i=1}^{24} $, with $ X^+= ( X^0+X^{25}),$ and $  X^-= ( X^0-X^{25})$. Then the unphysical degrees of freedom are removed by observing that 
 there are no oscillations in the $ X^+$ direction in In the light-cone gauge. Furthermore,   the  oscillation in $X^-$ direction can be expressed in terms of other string  oscillations, and need not be separately analyzed.  Thus, in the light-cone gauge, only the  string oscillations in the $\{  X^i\}_{i=1}^{24} $  direction have to be considered. The string algebra in these  directions can be written as   
\begin{equation}
[\hat{\alpha}_m^i ,\hat{\alpha}_n^j] =m\eta^{ij} \delta_{m+n,0}.
\end{equation} 
This can  be written as
$
[\hat{\alpha}_m^i ,\hat{\alpha}_{-m}^i] =m  $ for $ n =  -m $ and   $ (\hat{\alpha}_m^i)^\dagger =  \hat{\alpha}_{-m}^i
$ with  $\{i,j\}$,  taking  values in $\{1, 2,3,..,24\}$.

We define a vacuum as $ |0, p\rangle $ such that $\hat{p}^\mu|0, p\rangle = p^\mu|0, p k\rangle$. Using this vacuum, we can construct a state   $|k, 0\rangle$ from $\alpha_k|0, p\rangle $. 
We start by constructing  a string density matrix $\hat{\rho}$ from these string states as 
\begin{equation}
\hat{\rho}= \mathcal{N} \sum_l\sum_k d_k{d_l}^*|k;p\rangle\langle l;p|, 
\end{equation} 
$ \mathcal{N}=1/\sum_k |d_k|^2$where $\mathcal{N} $ is the normalization constant. 
This string density state evolves from $\hat{\rho} (0)$ at time $\tau =0$ to $\rho (t)$ at 
time $\tau = t$, such that 
\begin{equation}
   \hat{\rho}(t)=\hat{U}(t)\hat{\rho} (0) \hat{U}^{\dagger}(t). 
\end{equation}
Here $U$ is a suitable unitary operator, which evolves the initial string density state to the final string density state  \cite{Za,  5ab}. Such a unitary evolution occur due to the interaction of string states by a background field. Thus, for example, with a background  field \cite{y0, y4, y1, y2}.   The the  Polyakov action with a background  $B^{\mu\nu} (X)$ field,  gravitational  field with metric $G^{\mu\nu} (X)$, and dilaton field $\Phi (X)$ (which couples to the two-dimensional world-sheet Ricci scalar $R$), can be written as  
\begin{eqnarray}
     {S} =  \frac{1}{4 \pi \alpha'} \int d\sigma ^2 \Big[ \eta^{\alpha\beta}  \left(  G^{\mu\nu}(X) + i B^{\mu\nu} (X) \right)  \partial_{\alpha}X_{\mu}  \partial_{\beta} X_{\nu} + \alpha' \Phi (X) R \Big].
\end{eqnarray}
Here the background field  will interact with the string for a time $T$, where $T$ would be defined in the target space. So, we can perturb this string  with a background field, and write  the resulting total world-sheet Hamiltonian as 
\begin{equation}
    \hat{H} = \Tilde{H}_0+\Tilde{H}_I,
\end{equation}
where $\Tilde{H}_0 $ is the original free string Hamiltonian, and $\Tilde{H}_I  $
is the interaction of the string fields with the background  field. Here we have assumed that the   is time dependent, and we can separated the temporal dependence of $\Tilde{H}_I(\tau) $ into $ \lambda \chi (\tau) $. We can also extract a constant $\lambda$ from $\Tilde{H}_I  $, from the interaction Hamiltonian, which would act as an effective coupling constant, and  write 
    \begin{equation} 
    \Tilde{H}_I  = \lambda\chi (\tau) \int d\sigma  \mathcal{H}(\sigma).
    \end{equation} 
  Here we have chosen the gauge $\tau = T$, with $T$ being the time in target spacetime. With this choice of gauge, we can   view $\chi(\tau) $  as a switching function, which turns the background fields on at $\tau>0$ and off at $\tau < t$ on the world-sheet. Furthermore,  and $ \mathcal{H}(\sigma) $ will act as the   smearing function representing the interaction. Now the interaction is turned on at time $\tau >0$ and turned off at time $\tau = t$ such  that $\Tilde{H}(0)= \Tilde{H}_0$ and $\Tilde{H}(t)= \Tilde{H}_0$ as $\Tilde{H}_I(0) =\Tilde{H}_I(t) = 0 $.  So, using this interaction Hamiltonian $ \hat{H} $, we can write the unitary transformation $U$ as 
      \begin{equation}
      \hat{U} (t) = \mathcal{T} \exp{\left(-i  \lambda\chi (\tau) \int d\sigma  \mathcal{H}(\sigma)    \right)}, 
    \end{equation}
where $\mathcal{T}$ denotes time ordering between $0<\tau<t$.
    Using the Dyson expansion, we can express this unitary operator in terms of the a series as 
    \begin{equation}
    \hat{U}(\tau)=1+\hat{U}^1(\tau)+\hat{U}^2(\tau)+O(\lambda^3), 
    \end{equation}
    where $\hat{U}^1(\tau)$ and  $\hat{U}^2(\tau)$ are the first and second order terms in the Dyson expansion. 
    
We want to analyze the effect of such interaction on a quantity defined in string theory. This quantity can be a quantity like the square of the total  mass $M^2$   of the string or the angular momentum of the string $J$. Both these quantities are calculated from world-sheet dynamics as eigenvalues of operators defined on world-sheet, i.e. $\hat{M}^2$ and $\hat{J}$. However, they also  have important physical implications for particle physics, which can be observed in target space.   Thus, even though such quantities can be calculated from world-sheet, they have important physical significance in target space, which can be tested in particle physics experiments. We could develop the argument specifically for $M^2$ or $J$, but here we will keep our argument general, and define a general operator on world-sheet as $\hat{A}$. The argument can be applied to any such operator, which can be obtained from world-sheet dynamics, such as $\hat{M}^2$ or $\hat{J}$. 

As the string interacts with the background  field, we expect its oscillations to change. This would in turn change the value of quantities defined on the world-sheet, such as 
$\hat{M}^2$ or $\hat{J}$.  It would be important to measure the change in such values. Usually, such a    change would  be expected to be measured, by first making an measurement at $\tau = 0$, then making the string interact with a  field, and finally making another measurement at $\tau =t$. Even though this seems straightforward, it would in principle involve making two projective measurements on world-sheet. The problem here is that  two such projective measurements cannot be made on a system with an Lorentz structure due to problems like  superluminal signals  \cite{5cd, 5mn}, and world-sheet of string has an intrinsic Lorentz structure build into it. However, such   problems can be resolved for systems with a Lorentz structure by a scheme called the Ramsey scheme \cite{6ab, 6ba}. The Ramsey scheme has already been applied to quantum field theories \cite{5}. As the world-sheet of string theory is modeled by a conformal field theory, we propose that the change in the world-sheet operator $\hat{A}$ can be obtained by using a string theoretical generalization of the Ramsey scheme. 

\section{Characteristic Function }
We can construct the characteristic function for this process. 
As world-sheet has a Lorentz structure on it, we can use the  Ramsey scheme to obtain information related to two   world-sheet projective measurements.  So, if  $|A_i\rangle$  are the initial eigenvectors of the general world-sheet operator $|\tilde A_i\rangle$   with eigenvalues $\tilde A_i$. To obtain the information about change in the eigenvalues of  $\hat{A}$ due to its interaction with the background field, we will now generalize  Ramsey scheme \cite{6ab, 6ba}  to string theory. This will be done by first expressing the world-sheet operators using their eigenvectors and eigenvalues as    
  \begin{eqnarray}
    \hat{A}(0)=\sum_{i}A_i|{A}_i\rangle \langle {A}_i|,   \,\,\,\,\,
    \,\,\,\,\,\,\,\,\,\,\,\,\,\,\,\,\,\,\,\,\,\,\,\,\,
    \hat{A}(t)=\sum_{j}\tilde A_j|{\tilde A}_j\rangle \langle {\tilde A}_j|.
    \end{eqnarray}

We can write  the probability for measuring  eigenvalue $A_i$ at time $\tau =0$ as $p^0_i = |\langle {A}_i|\rho|{A}_i\rangle|^2$. The eigenvectors  of $\hat{A}$ evolves from $|A_i\rangle $ to $|\tilde A_l\rangle$ due to the interaction with the interaction with background field. As this evolution is captured by the unitary transformation $\hat{U}$, we  can also write   the  conditional probability for measuring eigenvalue $\tilde A_l$, at time $\tau = t$, if the initial eigenvalue was $A_i$ at time $\tau =0$ as $p^{\tau}_{l|i} = \langle \tilde {A}_j|U | {A}_i\rangle|^2 $. We can write  the difference between  these initial and the final eigenvalues of $\hat{A}$ as \begin{eqnarray}  
\Bar{A}_{i,j}= \tilde A_j-A_i.
\end{eqnarray}
Using $ p^0_i $ and $ p^{\tau}_{l|i}$, we obtain the  probability associated with the occurrence  of this difference $\Bar{A}^{i,l}$ as     
  \begin{eqnarray}
p_{i,j} &=& p^0_i p^{\tau}_{j|i}  \nonumber \\ 
&=& |\langle {A}_i|\rho|{A}_i\rangle|\langle \tilde {A}_j|U | {A}_i\rangle|^2.
\end{eqnarray}
As we have the probabilities associated with the difference $\Bar{A}_{i,l}$, we can write the full probability distribution corresponding corresponding to it. This can be done by defining a distribution variable       $\mathcal{A}$, which would correspond to different values of   $\Bar{A}_{i,l}$, if there was no degeneracies. However,  to account for degeneracies in  the eigenvalues  of $\hat{A}$, we define the  probability distribution as 
\begin{equation}
P(\mathcal{A})=\sum_{ij}p_{i,j}\delta(\mathcal{A}-\Bar{A}_{i,j}).
\end{equation}

%Now we are going to analyze a string theoretical process, such that the 
%initial and final string  states are free (perturbation is only turned on  between them)
%\begin{eqnarray}
%[\hat{\rho},\hat{A}]=0, \,\,\,\,\,\,\,\,\,\,\,\,\,\,\,\,\,\,\,\,\,\,  [\hat{A}, \hat{U}^{\dagger}\hat{A}U] =0
%\end{eqnarray}
As the probability distribution for the variable $\mathcal{A}$ can be written as $P(\mathcal{A})$, we can write the average value for $\mathcal{A}$ using   $P(\mathcal{A})$ as 
  \begin{eqnarray}
           \Bar{ \mathcal{A}}=\int \sum_{ij}p_{i,j}\delta(\mathcal{A}-\Bar{A}_{i,j})\mathcal{A}d\mathcal{A}.
     \end{eqnarray}
Using the expression for $p_{i,j}$ and integrating this expression, we obtain    
       \begin{eqnarray}
           \Bar{ \mathcal{A}}
         &=& \sum_{ij} |\langle {A}_i|\rho|{A}_i\rangle|\langle \tilde {A}_j|U | {A}_i\rangle|^2 \left[\tilde A_j (\tau) -A_i (0)\right]. 
         \end{eqnarray}
 Now using $[\hat{\rho},\hat{A}]=0, $ and $  [\hat{A}, \hat{U}^{\dagger}\hat{A} \hat{U}] =0$, we observe that
 \begin{equation} 
 \Bar{ \mathcal{A}} = \sum_l p_l (\tau) A_l{(\tau)}-\sum_i p_i (0) A_i{(0)},
 \end{equation}  
 where    $ p_l (\tau) = \sum_i p _{i,l} (\tau)$ and $p_i (0) = \sum_l p_{i,l} (0) $. 
We observe that $\sum_l p_l (\tau) A_l{(\tau)} = tr[\mathcal{A}{(\tau)}\rho{(\tau)}] $ and $\sum_i p_i (0) A_i{(0)}  = tr[\mathcal{A}{(0)}\rho{(0)}] $, and so  we can write  
\begin{equation}
  \Bar{ \mathcal{A}} = tr[\mathcal{A}{(\tau)}\rho{(\tau)}]-tr[\mathcal{A}{(0)}\rho{(0)}].
\end{equation}
 We can  define the corresponding characteristic function of  $\mathcal{A}$ as
  \begin{equation}
    \Tilde{P}(\mu)=\int P(\mathcal{A})e^{i\mu \mathcal{A}} d\mathcal{A}=\langle e^{i\mu \mathcal{A}}\rangle.
  \end{equation} 
Here  $\mu$ is  the parameter used in the Ramsey scheme. We first define an string auxiliary qubit, such that its ground state is   $|0\rangle$,  and its  excited state is   $|0\rangle$. This string auxiliary qubit is defined in such a way, that it is capable of transferring   information from a string density matrix. This can be done by first  
 coupling it  to the string density state $\rho (0)$. Initially both  the  string and qubit are  in the ground state. So, we  can write the    product state of the system as   
\begin{equation}
\hat{\rho}_\text{tot}=   \hat{\rho}\otimes\hat{\rho}_\text{aux}.
\end{equation} 
After coupling the  auxiliary qubit to the string density state, we apply a Hadamard operator to the qubit.   Now the string interacts with the background field, and this evolves the  total state of the  system by a  unitary   operator \cite{5} 
 \begin{equation}
     \hat{C} (\mu) = \hat{U} e^{-i\mu  \hat{H}(0) } \otimes|0\rangle\langle0|+ e^{-i\mu  \hat{H}(t)}\hat{U}\otimes|1\rangle\langle1|.
 \end{equation}
Using this unitary operator, the  string qubit state can  be expressed as \begin{equation}
    \hat{\rho}_\text{aux}= \text{Tr} _{X}[\hat{C}(\mu) \hat{\rho}_\text{tot} \hat{C}^\dagger (\mu)].
\end{equation}
Here we have defined a string trace operation   $\text{Tr}_{X}$, which   traces over the string states. After evolving the system, we apply  a final Hadamard operation to the qubit state, and extract the information about the evolution of the system by the background field. This is done as we now have an   expression for $\hat{\rho}_\text{aux}  $,   whose explicit  form will depend on the precise form of the  operator $\hat{A}$. This final expression for the string auxiliary qubit $\hat{\rho}_\text{aux}  $, can then be compared to the general expression for   $\hat{\rho}_\text{aux}   $ as  \cite{6ab, 6ba}  
\begin{equation}
\hat{\rho}_\text{aux}    =
\frac12\Big\{\mathbb{1} + \text{Re} [\tilde{P}(\mu )]\hat{\sigma}_z + \text{Im} [\tilde{P}(\mu )]\hat{\sigma}_y\Big\}~. 
\label{eq:rho_aux}
\end{equation}
This expression  is general, and will also hold for this string theoretical process. 
Thus, we can obtain $\tilde{P}(\mu )$ using a specific form of the string auxiliary qubit. So, 
this characteristic function is obtained from the string  Ramsey scheme. 
\section{Fisher Information}
We can use this characteristic function to analyze the behavior of physical quantities due to such an interaction with background field.  
It   can   be used to explicitly obtain the value of this average difference $\Bar{ \mathcal{A}}$ as 
 \begin{eqnarray}
  \Bar{ \mathcal{A}} =i  \frac{d}{d\mu}  
  \Tilde{P}(\mu).  
 \end{eqnarray}
  
So, basically we start from initial world-sheet string states. Then we perturb this system, and so this perturbation changes the string states. Then we obtain the difference between the initial and final eigenvalues of $\mathcal{A}$ using Ramsey scheme  \cite{6ab, 6ba}. Now if $\mathcal{A}$ does not contain any information about string states, then this average difference will vanish for such a string theoretical process, $\Bar{ \mathcal{A}} = 0$. 

The fisher information is used for  a quantity, which is not represented by a quantum 
operator \cite{1a, 2a, 1b, 2b}. As the difference of two quantities during a string  theoretical process cannot be represented by a quantum operator, we will use fisher information to analyze it. We can associate a parameter, say $\mu$ with the change in a physical quantity $\mathcal{A}$. Here $\mathcal{A}$ is a quantity defined on world-sheet such as  the change in the mass square or spin of strings during a string theoretical process. We can then use this parameter to obtain the fisher information associated with the change in that physical quantity, during a string theoretical process. If there is no fisher information, then the change in the quantity cannot have a physical meaning. However, if there is a change in the fisher information, then we can assume that the change in the physical quantity will have a physical meaning. We can write the fisher information associated with $\mathcal{A}$  as 
\begin{eqnarray}
   F =\int P(\mathcal{A}) \left| \frac{\partial }{\partial \mu }\log P(\mathcal{A})\right|^2 d\mathcal{A}.
\end{eqnarray}
To analyze string theoretical processes on world-sheet, we need to analyze the change of string states on the world-sheet. This change of string states can be measured by measuring the change in a quantity on the world-sheet, if that quantity contains information about string states of the world-sheet. Thus, we would start by defining a quantity on the world-sheet, represented by an operator $\mathcal{A}$, such that it contains information about world-sheet string states.  
However, if the quantum fisher information does not vanish (and so operator $\mathcal{A}$ contains information about string states), then 
we can analyze a string theoretical process by measuring the change in the value of  $\mathcal{A}$ during such a process.   

\section{Conclusion}
It is possible to obtain quantities, like mass and spin dynamically from the world-sheet dynamics of strings. It is also possible for strings to interact with background fields. 
In this letter, we  have investigated how such an interaction will change the a physical quantity. This has been done by  constructing  the characteristic function for such a string theoretical process.  We have argued  that such a    characteristic function cannot be constructed using the  usual two point measurements,  as the world-sheet has an intrinsic  Lorentz structure. We have also resolved this problem by  generalizing  the Ramsey scheme    to world-sheet. This has been done by coupling an string auxiliary qubit to a string density state. After the first Hadamard operation, this combined system is evolved due to the interaction of a background field.  Then a second Hadamard gate is applied, and the information about the process is extracted from the qubit. This is done without making two point measurements. It is used to obtain the    characteristic function for such a process.   Finally, using the  characteristic function,  we  also obtain   fisher information for the difference between the initial and final values of  a quantity, in a string theoretical process. 

It would be interesting to apply this formalism to specific operators defined on the world-sheet of string theory, such as $\hat{M}^2$ or $\hat{J}$. Then using a specific perturbation by a background field, we can explicitly obtain the characteristic function for such operators. This can then be used to analyze a change in their value, after the string has interacted with a background field. It would be interesting to analyze the consequence of such string theoretical processes. It is also possible to generalize this analysis to thermal states. This can be done by using KMS state instead of vacuum state.

\end{document}